\documentclass[twocolumn]{aastex631}
\usepackage[T1]{fontenc} 
\usepackage{amsmath}
\usepackage{amssymb}
\usepackage[version=4]{mhchem}
\usepackage{hyperref}

\newcommand{\e}{$_\oplus$}
\newcommand{\p}{$\pm$}
\newcommand{\ms}{m\,s$^{-1}$}

\newcommand{\gaia}{\textit{Gaia}}

\newcommand{\sol}{$_\odot$}
\newcommand{\ror}{$R_{\rm p}/R_\star$}
\newcommand{\tess}{\textit{TESS}}
\newcommand{\gcm}{g~cm$^{-3}$}

\begin{document}
\newcommand{\PSUAA}{Department of Astronomy \& Astrophysics, 525 Davey Laboratory, The Pennsylvania State University, University Park, PA, 16802, USA}
\newcommand{\PSUCEHW}{Center for Exoplanets \& Habitable Worlds, 525 Davey Laboratory, The Pennsylvania State University, University Park, PA 16802, USA}
\newcommand{\PSETI}{Penn State Extraterrestrial Intelligence Center, 525 Davey Laboratory, The Pennsylvania State University, University Park, PA 16802, USA}
\newcommand{\UA}{Steward Observatory, The University of Arizona, 933 N.\ Cherry Ave, Tucson, AZ 85721, USA}
\newcommand{\Caltech}{Department of Astronomy, California Institute of Technology, Pasadena, CA 91125, USA}
\newcommand{\JHU}{Department of Physics \& Astronomy, Bloomberg Center, Johns Hopkins University, Baltimore, MD 21218, USA}
\newcommand{\Macquarie}{School of Mathematical and Physical Sciences, Macquarie University, Balaclava Road, North Ryde, NSW 2109, Australia}
\newcommand{\CUBoulder}{Department of Physics, 390 UCB, University of Colorado, Boulder, CO 80309, USA}
\newcommand{\JPL}{Jet Propulsion Laboratory, California Institute of Technology, 4800 Oak Grove Drive, Pasadena, CA 91109, USA}
\newcommand{\MITEAPS}{Department of Earth, Atmospheric, and Planetary Sciences, Massachusetts Institute of Technology, Cambridge, MA 02139, USA}
\newcommand{\MITKavli}{Department of Physics and Kavli Institute for Astrophysics and Space Research, Massachusetts Institute of Technology, Cambridge, MA 02139, USA}
\newcommand{\UCI}{Department of Physics \& Astronomy, The University of California, Irvine, Irvine, CA 92697, USA}
\newcommand{\Carnegie}{Earth and Planets Laboratory, Carnegie Institution for Science, 5241 Broad Branch Road, NW, Washington, DC 20015, USA}
\newcommand{\PSUICS}{Institute for Computational and Data Sciences, The Pennsylvania State University, University Park, PA 16802, USA}
\newcommand{\PSUCASt}{Center for Astrostatistics, 525 Davey Laboratory, The Pennsylvania State University, University Park, PA 16802, USA}
\newcommand{\Princeton}{Department of Astrophysical Sciences, Princeton University, 4 Ivy Lane, Princeton, NJ 08540, USA}
\newcommand{\IAS}{Institute for Advance Study, 1 Einstein Drive, Princeton, NJ 08540, USA}
\newcommand{\Tsinghua}{Department of Astronomy, Tsinghua University, Beijing 100084, China}
\newcommand{\FlatironCCA}{Center for Computational Astrophysics, Flatiron Institute, 162 Fifth Avenue, New York, NY 10010, USA}
\newcommand{\ETH}{ETH Zurich, Institute for Particle Physics \& Astrophysics, Zurich, Switzerland}
\newcommand{\UCO}{UC Observatories, University of California, Santa Cruz, CA 95064, USA}
\newcommand{\SantaCruz}{University of California, Santa Cruz}
\newcommand{\WMKO}{W.\ M.\ Keck Observatory, 65-1120 Mamalahoa Hwy, Kamuela, HI 96743, USA}
\newcommand{\SSL}{Space Sciences Laboratory, University of California, Berkeley, CA 94720, USA}
\newcommand{\UH}{Institute for Astronomy, University of Hawai‘i, 2680 Woodlawn Drive, Honolulu, HI 96822, USA}
\newcommand{\UCB}{Department of Astronomy, 501 Campbell Hall, University of California, Berkeley, CA 94720, USA}
\newcommand{\UCLA}{Department of Physics \& Astronomy, University of California Los Angeles, Los Angeles, CA 90095, USA}
\newcommand{\nexsci}{NASA Exoplanet Science Institute/Caltech-IPAC, California Institute of Technology, Pasadena, CA
91125, USA}
\newcommand{\COO}{Caltech Optical Observatories, California Institute of Technology, Pasadena, CA 91125, USA}
\newcommand{\Sydney}{Sydney Institute for Astronomy (SIfA), School of Physics, University of Sydney, NSW 2006, Australia}
\newcommand{\Kansas}{Department of Physics and Astronomy, University of Kansas, Lawrence, KS, USA}
\newcommand{\Warwick}{Physics Department, University of Warwick, Coventry CV4 7AL, United Kingdom}
\newcommand{\Yale}{Department of Astronomy, Yale University, New Haven, CT 06520, USA}
\newcommand{\ICL}{Astrophysics Group, Department of Physics, Imperial College London, Prince Consort Rd, London SW7 2AZ, UK}
\newcommand{\Schmidt}{Astrophysics \& Space Institute, Schmidt Sciences, New York, NY 10011, USA}
\newcommand{\Amsterdam}{University of Amsterdam}
\newcommand{\ND}{Department of Physics and Astronomy, University of Notre Dame, Notre Dame, IN 46556, USA}
\newcommand{\Geneva}{Observatoire Astronomique de l'Université de Genève, Chemin Pegasi 51, 1290 Versoix, Switzerland}
\newcommand{\kansas}{Department of Physics \& Astronomy, University of Kansas, 1082 Malott,1251 Wescoe Hall Dr., Lawrence, KS 66045, USA}
\newcommand{\chicago}{Department of the Geophysical Sciences, University of Chicago, 5734 South Ellis Ave., Chicago, IL 60637, USA}
\newcommand{\DGPSCaltech}{Division of Geological and Planetary Sciences,
1200 E California Blvd, Pasadena, CA, 91125, USA}
\newcommand{\GSFC}{NASA's Goddard Space Flight Center, Greenbelt, MD 20771, USA}
\newcommand{\IFA}{Institute for Astronomy, University of Hawai`i, 2680 Woodlawn Drive, Honolulu, HI 96822, USA}

\title{Revisiting the Exo-Mercury Candidate GJ~367~b with ESPRESSO and a Self-Consistent Tidal Distortion Model}

\author[0000-0001-7058-4134]{Rena A.~Lee}
\altaffiliation{NSF Graduate Research Fellow}
\affiliation{\IFA}

\author[0000-0002-8958-0683]{Fei Dai}
\affiliation{\IFA}
\affiliation{\DGPSCaltech}
\affiliation{\Caltech} 

\author[0000-0002-3286-3543]{Ellen M.~Price}
\affiliation{Department of Astronomy, University of Chicago, 5640 South Ellis Ave., Chicago, IL 60637, USA} 

\author[0000-0002-7127-7643]{Te Han}
\affil{\UCI} 

\author[0000-0001-8627-9628]{Davide Gandolfi}
\affiliation{Dipartimento di Fisica, Università degli Studi di Torino, Via Pietro Giuria, 1, 10125 Torino, Italy}

\author[0000-0002-6532-4378]{Mathias Zechmeister} 
\affiliation{Institut f\"ur Astrophysik und Geophysik, Georg-August-Universit\"at  G\"ottingen, Friedrich-Hund-Platz 1, 37077 G\"ottingen, Germany} 

\author[0000-0001-7409-5688]{Guðmundur Stef\'ansson}
\affiliation{Anton Pannekoek Institute for Astronomy, University of Amsterdam, Science Park 904, 1098 XH Amsterdam, The Netherlands} 

\author[0000-0002-3610-6953]{Jiayin Dong}
\affiliation{Department of Astronomy, University of Illinois at Urbana-Champaign, Urbana, IL 61801, USA}
\affiliation{Center for Astrophysical Surveys, National Center for Supercomputing Applications, Urbana, IL 61801, USA} 

\author[0000-0003-1762-8235]{Simon H.\ Albrecht}
\affiliation{Stellar Astrophysics Centre, Department of Physics and Astronomy, Aarhus University, Ny Munkegade 120, DK-8000 Aarhus C, Denmark} 

\author[0000-0002-9910-6088]{Kristine W.~F.\ Lam}
\affiliation{Centre for Astronomy and Astrophysics, Technical University Berlin, 10585 Berlin, Germany}
\affiliation{Institute of Planetary Research, German Aerospace Center, 12489 Berlin, Germany} 


\author[0000-0001-5180-2271]{Federica Chiti}
\affiliation{\IFA} 

\author[0000-0002-4284-8638]{Jennifer L.\ van Saders}
\affiliation{\IFA} 

\author[0000-0001-8832-4488]{Daniel Huber}
\affiliation{\IFA} 

\author[0000-0002-5375-4725]{Heather A. Knutson}
\affiliation{\DGPSCaltech} 

\author[0000-0001-6588-9574]{Karen A.\ Collins}
\affiliation{Center for Astrophysics \textbar \ Harvard \& Smithsonian, 60 Garden Street, Cambridge, MA 02138, USA} 

\author[0000-0002-0659-1783]{Michael Zhang}
\altaffiliation{Heising-Simons Foundation 51 Pegasi b Postdoctoral Fellow}
\affiliation{Department of Astronomy \& Astrophysics, University of Chicago, 5640 S Ellis Avenue, Chicago, IL 60637, USA} 

\author[0000-0003-0638-3455]{Leslie A.~Rogers}
\affiliation{Department of Astronomy \& Astrophysics, University of Chicago, 5640 S Ellis Avenue, Chicago, IL 60637, USA} 

\author{Eleonora Armano}
\affiliation{Dipartimento di Fisica, Università degli Studi di Torino, Via Pietro Giuria, 1, 10125 Torino, Italy}

\author[0000-0002-4480-310X]{Casey L. Brinkman}
\altaffiliation{Trottier Space Institute Fellow}
\affiliation{McGill University, Trottier Space Institute, 3550 rue University, Montreal, QC H3A 2A7, Canada} 

\author[0000-0003-2657-3889]{Nicholas Saunders}
\altaffiliation{YCAA Prize Postdoctoral Fellow}
\affiliation{\IFA}
\affiliation{\Yale} 

\author[0000-0003-3244-5357]{Daniel Hey}
\affiliation{\IFA} 

\begin{abstract}
    We report revised mass and radius measurements for GJ~367~b, an ultra-short-period (7.7 hr) sub-Earth in a multi-planet system orbiting a nearby ($\sim$9 pc) M dwarf host. Previous mass and radius measurements have suggested GJ~367~b has an anomalously high bulk density, close to that of solid iron. The existence of such an iron-rich planet is in tension with established planet formation scenarios. We utilized newly available \tess\ short-cadence photometry to constrain the radius of GJ~367~b to 0.736 \p\ 0.035 R\e. We consider observational and modeling effects such as photometric dilution, stellar activity, and tidal distortion to account for possible inaccuracies in the star and planet radius measurements. From our radial velocity (RV) analysis using VLT/ESPRESSO data covering nearly the full orbit in a single night, we find a mass of 0.503 \p\ 0.078 M\e, corresponding to a bulk density of 6.9$^{+1.6}_{-1.4}$ \gcm. We present a new tidal distortion and interior composition modeling framework to assess the iron mass fraction of GJ~367~b. Considering several different interior composition assumptions and radial aspect ratios, we find an iron fraction of $\sim$50-70\%, which is broadly consistent with that of Mercury and not as iron rich as previously suggested.
\end{abstract}

\section{Introduction} \label{sec:intro}

GJ~367~b, an ultra-short-period (USP; $P_{\rm orb}<1$ day) sub-Earth around an M dwarf host, was first discovered by \citet{Lam2021} with an orbital period of 7.7 hours, a radius of $0.718\pm0.054~R_{\oplus}$, and a mass of $0.546\pm0.078~M_{\oplus}$, making it one of the smallest and densest exoplanets characterized to date. Its bulk density of $\sim8$\gcm\ \citep{Lam2021} places it firmly in the “super-Mercury” regime, requiring an iron-rich interior with a core mass fraction comparable to or exceeding that of Mercury. Follow-up photometric and radial velocity studies have refined these parameters and strengthened the evidence that GJ~367~b is a rocky, metal-enriched planet \citep[e.g.,][]{Adibekyan2021,zhang_gj_2024}, with recent work by \citet{goffo_company_2023} suggesting that GJ~367~b may even be denser than a pure-iron composition ($M=0.633\pm0.050$ M\e, $R=0.699\pm0.024$ R\e, and $\rho=10.2\pm1.3$ \gcm). Such a result challenges our understanding of rocky planet formation: possible scenarios—including mantle-stripping giant impacts \citep{Benz2007,Asphaug2014}, silicate evaporation from intense irradiation \citep{Kite2016,Ito2022}, and disk-driven chemical fractionation \citep{ElkinsTanton2008,Dorn2019}—may not reproduce planets of such high core mass fraction at a significant frequency. \citet{goffo_company_2023} emphasized that, while some of these formation scenarios can account for such a high-density planet, the specific formation pathway which resulted in the planet's formation remains uncertain. While dynamical simulations \citep[e.g.,][]{Scora2024,Tajer2025} demonstrate that Mercury analogues can form naturally from excited inner-disk conditions, even these models rarely produce densities as extreme as inferred for GJ~367~b. This tension suggests that GJ~367~b either followed an unusual evolutionary pathway—such as extreme mantle-stripping collisions \citep{Reinhardt2022}, collisional erosion in dynamically perturbed disks \citep{Scora2024}, or preferential silicate loss under intense stellar irradiation \citep{Wurm2013,Kite2016}—or that our current theoretical frameworks for terrestrial planet formation and differentiation require revision. Conversely, the apparently anomalous density of GJ~367~b may reflect inaccuracies in its measured radius, mass, or tidally distorted structure, as has recently been demonstrated for other putatively iron-rich planets \citep[e.g., HD~93963~A~b, Kepler-100~b, and Kepler-106~b and e;][]{Brinkman2025,romy2023,Weiss2024}.

Beyond its implications for formation theory, GJ~367~b exemplifies why M-dwarf planetary systems have become central to the study of terrestrial exoplanets. Due to their frequent transits, high planet-to-star radius ratios, and favorable spectroscopic contrasts, close-in planets orbiting M dwarfs are leading targets for detailed studies of rocky planet surfaces and atmospheres \citep[e.g.,][]{NutzmanCharbonneau2008,Dressing2013,deWit2016,Morley2017,Lustig-Yaeger2019,kreidberg_absence_2019}. The first phase curves and eclipse measurements of USPs such as 55~Cnc~e \citep{Zieba2022}, K2-141~b \citep{Barragan2018,Malavolta2018}, and LHS~3844~b \citep{kreidberg_absence_2019} have already revealed extreme dayside temperatures, possible magma-ocean surfaces, and the absence of substantial atmospheres. Accurate determinations of planetary masses, radii, and bulk compositions are critical for interpreting such observations from JWST and future facilities, where degeneracies between interior composition, surface gravity, and atmospheric scale height otherwise hinder robust characterization. Earth-sized and smaller planets remain an underrepresented subset of confirmed exoplanets, but they are a rapidly growing population of high-profile targets that hold key insights into the formation and evolution of terrestrial worlds under diverse stellar conditions. At the same time, the accurate characterization of these planets is complicated by stellar activity \citep[e.g.,][]{Newton2016,Morales2010}, magnetic radius inflation \citep{Somers2020,Kochukhov2021}, and the potential for systematic errors in radius and mass measurements \citep{Mann2015,Bouma2022}—all of which are particularly acute for planets orbiting low-mass stars like GJ~367.

In this work, we present revised planetary parameters for the sub-Earth USP GJ~367~b, and we examine the modeling and observational uncertainties that impede its accurate characterization. Section~\ref{sec:star} describes the stellar characterization of the host star, GJ~367. Section~\ref{sec:phot} details our transit modeling and radius determination for GJ~367~b, while Section~\ref{sec:rv} presents our radial velocity observations and mass measurement. We detail our self-consistent tidal distortion and interior composition model in Section~\ref{ssec:distortion}. We discuss the implications of stellar and planetary modeling uncertainties, as well as possible formation and evolution scenarios, in Section~\ref{sec:discussion}, and conclude with a summary of our results in Section~\ref{sec:sum}.

\section{Host Star Characterization}\label{sec:star}
 Extracting stellar parameters for M dwarfs is notoriously challenging; their spectra are plagued with ambiguous continua and blended atomic and molecular lines, and stellar evolution models systematically fail to model and explain observed scatter in physical relations such as mass-radius \citep{Kraus2011,Parsons2018} and radius-temperature \citep{Boyajian2012,Mann2015}. We employed {\tt isoclassify} \citep{huber2017,berger2020,berger2023} to derive stellar parameters for GJ~367. We used the {\tt isoclassify} direct method which uses extinction maps and bolometric corrections interpolated from the MIST evolutionary models \citep[][]{MIST} to derive stellar parameters based on the physical relations of \citet{mann2019}. Because GJ~367 is very nearby ($\sim$9 pc), we set the extinction to zero. We adopted 2MASS photometry, \gaia\ DR3 astrometry and photometry, as well as spectroscopic parameters ($T_{\rm eff}$, $\log g$, [Fe/H]) reported in \citet{goffo_company_2023} as inputs. For comparison, we also estimated the stellar mass and radius directly using the empirical scaling relations of \citet{mann2019} and found consistent results within 1$\sigma$ uncertainties. We additionally examined the stellar parameters derived from {\tt isoclassify} using the grid method, which integrates isochrone grids to determine posterior distributions of stellar parameters. The grid modeling results varied widely between different models (MESA and PARSEC), so we elected to adopt the directly derived parameters, which are listed in Table \ref{tab:params}. Our adopted stellar parameters are consistent with those reported in \citet{Lam2021} and \citet{goffo_company_2023}. 

 \begin{figure*}[!ht]
    \centering
    \includegraphics[width=\linewidth]{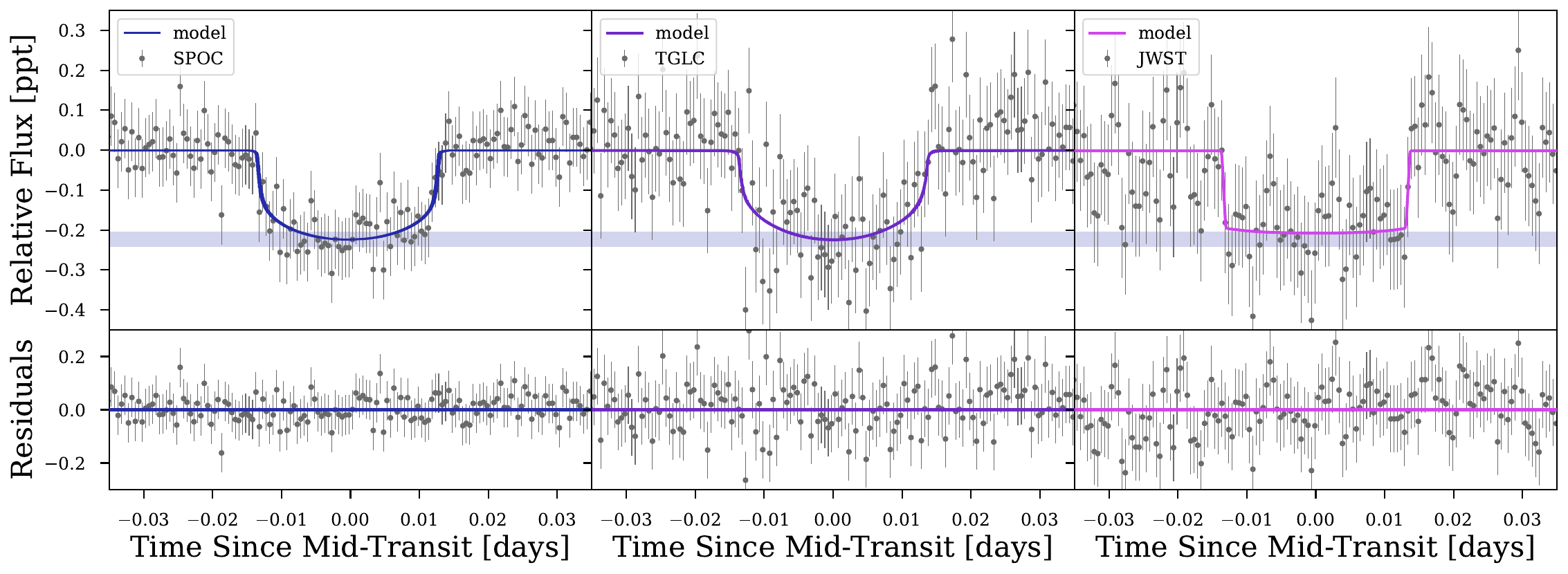}
    \caption{From left to right: SPOC PDCSAP, TGLC, and JWST photometry. \textbf{Top}: Binned (30-sec) and phase-folded light curve of GJ~367 confirming the transit signal of GJ~367~b (grey points). The best-fit transit model corresponding to each dataset (see Sec.\ \ref{ssec:tran}) is overplotted. The blue shaded region in each panel is the adopted transit depth and 1$\sigma$ uncertainty from the SPOC photometry fit, showing the transit depths are consistent. \textbf{Bottom}: Data minus model residuals. }
    \label{fig:transit}
\end{figure*}

\begin{deluxetable*}{lccc}
\tablecaption{GJ~367 system parameters. For values fit and/or derived in this work, we present the Posterior Median and 68.3\% Confidence Interval} 
\label{tab:params}
\tablehead{
\colhead{Parameter} & \colhead{Symbol} & \colhead{Prior} & \colhead{Value}}
\startdata
\textbf{Stellar}\\
TIC ID $^\dagger$  & ... & ...& 34068865  \\
R.A. (epoch 2015.5) $^\dagger$  & $\alpha$ &... & 09:44:29.15\\
Dec. (epoch 2015.5) $^\dagger$  & $\delta$ &... & -45:46:44.46\\ 
$V$ (mag) $^\dagger$ &  ...& ...&	$10.153 \pm 0.044$ \\
$K_s$ (mag) $^\dagger$ &... &... & $5.78\pm 0.02$\\
Effective Temperature (K)& $T_{\text{eff}}$ &... & 3535 $ \pm $ 127\\
Stellar Mass$~(M_{\odot})$ & $M_{\star} $ &... & 0.451 $ \pm $ 0.019\\
Stellar Radius$~(R_{\odot})$ $^\S$ & $R_{\star}$ & ...& 0.457 $ \pm $ 0.015\\
Stellar Density ($\rho_\odot$) & $\rho_\star$  & ...& 4.72 $ \pm $ 0.50\\
Limb Darkening q$_1$ \citep{exoplanet:kipping13} & ... & $\mathcal{U}$[0, 1] & 0.45 $ \pm $ 0.31\\
Limb Darkening q$_2$ \citep{exoplanet:kipping13} &  ... & $\mathcal{U}$[0, 1] & 0.22$ \pm $0.36\\
Parallax (mas) $^\ddagger$ & $\pi$ & ...& $106.1727 \pm 0.0141$\\
Distance (pc) $^\ddagger$ & \textit{d} &  ...& $9.4186 \pm 0.0012$ \\
\hline
\textbf{Planetary}\\
Equilibrium Temperature$^\ast$ & $T_{\text{eq}}$ & ... & 1329 $ \pm $ 50\\
Time of Conjunction (BJD-2457000) & $T_c$  &  $\mathcal{N}$[1544.13635, 10] & 1543.81414 $ \pm $ 0.00177 \\
Impact Parameter & $b$  & $\mathcal{U}$[0, 1+R$_{\rm p}$/R$_{\star}$] & 0.68 $\pm$ 0.04\\
Scaled Semi-major Axis & $a/R_\star$  & ... &  3.33 $ \pm $ 0.17 \\
Orbital Inclination (deg) & $i$  & ... & 78.6 $ \pm $ 0.7 \\
Orbital Eccentricity  & $e$  & 0 (fixed)& 0  \\
Orbital Period (days) & $P_{\rm orb}$   & $\mathcal{N}$[0.3219224, 10] & 0.3219226 $ \pm $ 0.000001\\
Planet/Star Radius Ratio & \ror\  & $\mathcal{N}\,\propto$ transit depth & 0.0148 $ \pm $ 0.0004\\
Planetary Radius ($R_\oplus$) $^\S$ & $R_{\rm p}$ & ... & 0.736 $ \pm $ 0.035\\
Planetary Mass ($M_\oplus$)  & $M_{\rm p}$ & ... & 0.503 $ \pm $ 0.078\\
RV Semi-amplitude (\ms)  & $K$ & \textit{Jeffreys}[0.1, 30]  & 0.78 $ \pm $ 0.12 \\
\hline
\textbf{GP Hyperparameters - Single Keplerian Fit} \\
ESPRESSO RV Jitter (\ms)  & $\sigma_{\rm jit,ESPRESSO}$ & \textit{Jeffreys}[0.1, 10] & $0.16^{+0.08}_{-0.04}$\\
HARPS RV Jitter (\ms)  & $\sigma_{\rm jit,HARPS}$ & \textit{Jeffreys}[0.1, 10] & $1.09^{+0.07}_{-0.06}$\\
GP variability amplitude (\ms) & $\eta_1$ & \textit{Jeffreys}[0.1,100] & $2.96^{+0.32}_{-0.27}$  \\
GP non-periodic characteristic length (days) & $\eta_2$ & $\mathcal{N}$[5, 25] & $4.52^{+0.46}_{-0.42}$\\
GP variability period (days) & $\eta_3$ & $\mathcal{U}$[1,200] & $60.00^{+0.66}_{-0.64}$ \\
GP periodic characteristic length (days) & $\eta_4$ &  $\mathcal{U}$[1,200] & $50.37^{+9.83}_{-9.97}$ \\
\enddata
\tablecomments{$^\dagger$TICv8.2; $^\ddagger$\textit{Gaia} DR3; $^\S$caveats summarized in Table \ref{tab:bias} and Sec. \ref{sec:discussion}; $^\ast T_{\text{eq}}$ assumes a low albedo of A$_B$=0.1}
\end{deluxetable*}

\section{Photometric Analysis}\label{sec:phot}
\citet{Lam2021} fit the transit signal of GJ~367~b using the \tess\ 2-min cadence light curve taken in Sector 9 (28 February to 26 March 2019), extracted using the Science Processing Operations Center (SPOC) pipeline \citep{jenkins_tess_2016}. Two years later, \citet{goffo_company_2023} combined the Sector 9 data with Sectors 35 and 36 (9 February to 2 April 2021) 20-sec cadence Presearch Data Conditioning Simple Aperture Photometry \citep[PDC-SAP;][]{Smith2012,Stumpe2012,Stumpe2014} light curves from SPOC to fit the transit of GJ~367~b. We present an independent transit fit using newly available sectors of \tess\ 20-second cadence data (Sec.\ \ref{ssec:tess}), a dilution assessment using {\tt TGLC} (Sec. \ref{ssec:dilution}), and a custom transit fitting routine to model the transit signal of GJ~367~b. 

\subsection{\tess\ Photometry}\label{ssec:tess}
 Since the previous analyses, \tess\ has observed GJ~367 in Sectors 63, 89, and 90 (10 March to 6 April 2023, 11 February to 9 April 2025), doubling the amount of photometric data available for transit analysis in the present work.\footnote{GJ~367 was also observed in Sector 62, but fell too close to the edge of the detector to generate reliable light curves.} We retrieved the SPOC PDC-SAP light curves from Sectors 9, 35, 36, 63, 89, and 90 available on the Mikulski Archive for Space Telescopes (MAST\footnote{\url{https://mast.stsci.edu/portal/Mashup/Clients/Mast/Portal.html}}) using {\tt lightkurve} \citep{lightkurve}. 20-sec cadence photometry was collected for each sector, excluding Sector 9, for which we collected the 2-min cadence photometry.

 As an independent check for possible dilution in the SPOC light curves, we generated \tess-\gaia\ Light Curves \citep[{\tt TGLC};][]{han_tessgaia_2023} to compare the photometric contamination factors in the two pipelines. {\tt TGLC} uses PSF forward-modeling and the \gaia\ nearby star catalog to provide a careful treatment of photometric contamination from nearby stars. The TGLC light curves were generated using the shortest-cadence Full Frame Images (FFIs) available in each sector, which ranged from 200-sec to 30-min. See Sec.\ \ref{ssec:dilution} for a further discussion on TGLC and dilution assessment.

\subsection{Transit Modeling}\label{ssec:tran}

We detrended the PDC-SAP light curves (Sec. \ref{ssec:tess}) to remove stellar and instrumental systematics using the {\tt wōtan} Python package \citep{hippke_wotan_2019}. After masking the transit signal of GJ~367~b based on the transit parameters reported in \citet{goffo_company_2023}, we applied an iterative sigma-clipping spline fit ({\tt wōtan/rspline}) to detrend the light curve, adopting a spline width of 0.8 days. This spline width is long enough to preserve the short ($\sim$0.3 day) planetary signal while still effectively removing stellar and instrumental variability on $<$1 day timescales. 

We constructed our transit model in the {\tt exoplanet} package \citep{foreman-mackey_exoplanet_2021}, a gradient-based probabilistic inference toolkit for modeling time-series astronomical data. To initialize the model, we used the transit parameters reported in \citet{goffo_company_2023} (orbital period, epoch, transit depth, and transit duration) as priors with uniform or broad normal distributions, in order to allow for an unconstrained search of the parameter space. We added Gaussian priors on the stellar radius and density which were derived in this work (see Sec.\ \ref{sec:star} and Table \ref{tab:params}). The {\tt exoplanet} toolkit employs the {\tt starry} package \citep{starry} to model the quadratic limb darkening parameters ($q_1,\;q_2$), following the formulation of \citet{exoplanet:kipping13}. Previous works \citep[e.g.,][]{goffo_company_2023,zhang_gj_2024} report eccentricities consistent with zero. This is expected for USPs due to rapid tidal circularization. We assume a circular orbit ($e=0$) for model simplicity. We also fit for the transit depth to derive the planet-to-star radius ratio (\ror), and the orbital inclination ($i$) derived from the impact parameter ($b$). We fit all transits with a single transit ephemeris.

We determined an initial maximum \textit{a posteriori} (MAP) solution from a nonlinear optimization using {\tt pyMC3} \citep{exoplanet:pymc3}. We set the MAP solution as the initial condition for sampling, and explored the parameter space with No U-Turn Sampling \citep[NUTS;][]{Hoffman2011} using a gradient-based Hamiltonian Monte Carlo \citep[HMC;][]{neal2011,betancourt2017} to determine the posterior distributions. The sampler was run with two chains for 1000 tune and 10,000 draw iterations (22,000 draws total). We adopt the median values of the posterior distributions for the transit parameters as our best-fit solutions. These values are listed in Table \ref{tab:params} along with the 68.3\% confidence intervals. The binned and phase-folded PDC-SAP light curve of GJ~367 is shown in the left panel of Fig. \ref{fig:transit} along with the best-fit transit model. Our measured transit radius of 0.736$\pm$0.035 R\e\ is consistent within 1$\sigma$ with that of \citet{goffo_company_2023} (0.699$\pm$0.024 R\e) and \citet{Lam2021} (0.718 $\pm$ 0.05 R\e). The slight tensions between the multiple transit radius determinations may reflect differences in the detrending treatment between studies. While \citet{goffo_company_2023} fit a second-order polynomial to the out-of-transit data in 2.5-hr windows about each transit midpoint, \citet{Lam2021} implemented a Savitsky-Golay filter on the PDCSAP light curve, as opposed to our masked-transit spline fit. 

\subsection{Addressing Possible Transit Dilution} \label{ssec:dilution}
Due to the large pixel size of \tess\ ($\sim21^{\prime\prime}$), photometric contamination from nearby stars can dilute planetary transit signals, leading to underestimated planetary radii if not properly corrected \citep{Han2025}. This effect is particularly relevant for small planets like GJ~367~b, where the transit depth is shallow and highly sensitive to even modest levels of contamination. We searched the \textit{Gaia} DR3 catalog for nearby bright sources within a radius equivalent to the \tess\ pixel size that may contribute to dilution. We identified one $\sim$10 mag fainter source, separated by $\sim5^{\prime\prime}$, that falls within the \tess\ target pixel file cutout. To assess the impact of dilution, we compared the \tess\ SPOC light curves with the TGLC light curves. Both data sets were detrended and fit with the same HMC transit parameter model to ensure consistency. From the TGLC light curves, we derived a planet-to-star radius ratio of \ror$=0.0143 \pm 0.0007$, which agrees with the SPOC measurement (\ror$=0.0148\pm0.0004$) within 1$\sigma$. We further validated this consistency by fitting the transit signal detected in JWST/MIRI photometry \citep{zhang_gj_2024}, which provides a higher-precision, independent check. From the reduced MIRI light curve available in \citet{zhang2024} we find \ror$=0.0135\pm0.001$. The phase-folded transit photometry and best-fit models from each dataset are shown in the middle and right panels of Fig.\ref{fig:transit}. Importantly, such a JWST-based validation of transit depth is rarely available for sub-Earth exoplanets, making GJ~367~b one of the few systems where photometric dilution can be ruled out with confidence. 

The degree of transit dilution can vary among different \tess\ sectors due to changes in spacecraft orientation and the placement of the target on the detector. For GJ367, we find no evidence of severe contamination across sectors (Fig.~\ref{fig:contamination}). Both the SPOC and TGLC pipelines estimate contamination factors of $\sim$12\% for their respective apertures, and both incorporate these corrections into their light curve processing. Although their contamination factors are not strictly comparable—since each pipeline defines its aperture differently—the fact that the corrected transit depths and inferred radii are in excellent agreement (Fig.\ref{fig:transit}) provides strong evidence that dilution is well accounted for. This cross-validation between pipelines increases confidence that the radius of GJ~367~b is not significantly biased by photometric contamination in the \tess\ data.

For our analysis, we adopt the SPOC result as the baseline measurement, since it achieves higher precision owing to its shorter cadence (20-second) photometry, which better samples the extremely short transit ingress and egress timescales of ultra-short-period planets like GJ~367~b \citep[$\lesssim 3$ min;][]{winn_exoplanet_2010}. The agreement between SPOC, TGLC, and JWST results suggests that dilution corrections in the \tess\ pipelines are robust, and that any residual contamination has a negligible effect on the inferred planetary radius. We therefore consider transit dilution to be a subdominant source of systematic error in this system, and we focus on other sources of bias---such as stellar radius inflation and spot contamination---in Sec.~\ref{sec:discussion}.

\begin{deluxetable}{DCC}
\tablewidth{0pt}
\tablecaption{VLT/ESPRESSO RV measurements\label{tab:rv}} 
\tablecolumns{3}
\tablehead{
\multicolumn2c{Time (BJD)} & \colhead{RV (\ms)} & \colhead{Uncertainty (\ms)}}
\decimals
\startdata
2460320.74873704 & -63.2 & 0.36 \\
2460320.77847757 & -63.88 & 0.42 \\
2460320.6310255 & -65.28 & 0.49 \\
2460320.62359065 & -64.57 & 0.51 \\
2460320.85965352 & -64.39 & 0.35 \\
2460320.77073292 & -63.43 & 0.39 \\
2460320.74144023 & -63.72 & 0.37 \\
2460320.67530257 & -63.8 & 0.4 \\
2460320.70469685 & -63.3 & 0.35 \\
2460320.82231352 & -63.6 & 0.36 \\
2460320.81506578 & -63.98 & 0.36 \\
2460320.69717679 & -64.15 & 0.33 \\
2460320.6679126 & -64.35 & 0.39 \\
2460320.8370245 & -64.37 & 0.34 \\
2460320.84441789 & -64.76 & 0.37 \\
2460320.66051412 & -64.76 & 0.43 \\
2460320.80761689 & -64.14 & 0.35 \\
2460320.71919471 & -63.59 & 0.38 \\
2460320.68991575 & -63.47 & 0.37 \\
2460320.68252993 & -63.86 & 0.35 \\
2460320.71196115 & -62.83 & 0.35 \\
2460320.80017566 & -64.34 & 0.34 \\
2460320.65311247 & -64.1 & 0.44 \\
2460320.85179666 & -63.7 & 0.36 \\
2460320.829715 & -64.14 & 0.35 \\
2460320.79288078 & -64.01 & 0.33 \\
2460320.73406407 & -63.26 & 0.36 \\
2460320.76329255 & -63.57 & 0.35 \\
2460320.64551112 & -64.27 & 0.47 \\
2460320.63824604 & -64.94 & 0.45 \\
2460320.75612485 & -62.93 & 0.34 \\
2460320.72667493 & -63.48 & 0.36 \\
2460320.78560838 & -63.13 & 0.35 \\
\enddata
\tablecomments{RV measurements are median-offset}
\end{deluxetable}

\section{Radial Velocity Analysis}\label{sec:rv}

\citet{goffo_company_2023} collected a total of 371 RVs (including the 77 RVs published in \citealt{Lam2021}) of GJ~367 with the High Accuracy Radial
velocity Planet Searcher \citep[HARPS;][]{Mayor2003} spectrograph on the ESO 3.6-m telescope at La Silla Observatory, Chile.  We fit the planetary RV modulation using both the combined HARPS and ESPRESSO datasets, and the ESPRESSO data alone (described below), to assess possible stellar activity and instrumental noise biasing the HARPS data. 

\subsection{ESPRESSO Observations}\label{ssec:espresso}

We obtained 33 RV measurements of GJ~367 with the Echelle SPectrograph for Rocky Exoplanets and Stable Spectroscopic Observations \citep[ESPRESSO;][]{Pepe2021} on the European Southern Observatory (ESO) Very Large Telescope (VLT) at Paranal Observatory, Chile. ESPRESSO is a state-of-the-art fiber-fed spectrograph that has achieved an unprecedented precision of $<$25 c\ms\ in RV measurements \citep{Pepe2021}. The observations were collected on 11 January 2024, with a baseline of 5.5 hr, just short than the full orbital period (7.7 hr) of GJ~367~b. This observation strategy helps to mitigate long-term stellar and instrumental correlated noise affecting the sensitive ($<$ 1 \ms) planetary RV signal. We set up ESPRESSO in the single Unit Telescope (1UT) configuration and used the high-resolution (HR) mode ($R\sim140,000$; 380-780 nm), with simultaneous calibration lamp observations. Each exposure was 600 sec, with an average S/N of $\sim$100 across all exposures. The spectra were reduced with the ESPRESSO data reduction software (DRS)\footnote{\url{https://ftp.eso.org/pub/dfs/pipelines/instruments/espresso/espdr-kit-3.3.12-1.tar.gz}}. 

For M dwarfs such as GJ~367, the template matching method typically is better suited for producing precise RVs compared to the cross-correlation function (CCF) technique used in the DRS. However, \citet{Silva2025} have elucidated a systematic trend in single-night RV measurements extracted via template matching, induced by microtellurics contaminating stellar templates constructed from single-night datasets. Thus, we elected to use the CCF RVs extracted using the DRS for our analyses, despite the larger individual measurement uncertainties and RMS scatter. The ESPRESSO RV data is given in Table \ref{tab:rv}. The HARPS RVs from previous literature \citep{Lam2021,goffo_company_2023}, extracted using a template matching code (NAIRA; \citealt{astudillo2017}), are not affected by this systematic, because the observations were taken and stellar templates were constructed over many nights. 

\subsection{ESPRESSO Data: Keplerian Orbital Fit}

 Because the ESPRESSO RVs were collected in a single night, we expect the data to be less prone to correlated noise caused by long-term stellar variability or instrumental drift compared to the HARPS dataset. Other short-term stellar light modulation, such as pulsations and granulation, have very low amplitudes and short periods in M dwarfs, and therefore should effectively be averaged out between exposures \citep{Kjeldsen1995,Chaplin2019}. Furthermore, the ESPRESSO instrument achieves a largely improved RV precision compared to HARPS. Because of the advantages the ESPRESSO dataset offers, we first elected to fit these data alone.

 We used the open source {\tt radvel} Python package \citep{fulton_radvel_2018} to fit the RV time series of GJ~367 with a Keplerian orbit. The model basis included these 5 orbital elements: the orbital period ($P_{\rm orb}$), time of conjunction ($T_c$), RV semiamplitude $K$, eccentricity $e$, and argument of pericenter $\omega$. We parameterized $e$ and $\omega$ jointly as $\sqrt{e}\cos{\omega}$ and $\sqrt{e}\sin{\omega}$. As in the transit fit, we again assumed a circular orbit and set these to zero. We additionally included the ESPRESSO white noise jitter term $\sigma_{ ESPRESSO}$, an arbitrary RV offset $\gamma$, and a linear acceleration $\dot\gamma$ in our model. We imposed Gaussian priors on $P_{\rm orb}$ and $T_c$ based on our transit analysis (Sec.\ \ref{ssec:tran}), along with Jeffreys priors on $K$ and $\sigma_{\rm ESPRESSO}$. A uniform prior was imposed on $\gamma$. We did not model the RV modulation of the other known planets, GJ~367~c and d, since RV modulation should be subsumed by $\gamma$ and $\dot\gamma$.

\begin{figure*}[!ht]
    \centering
    \includegraphics[width=\linewidth]{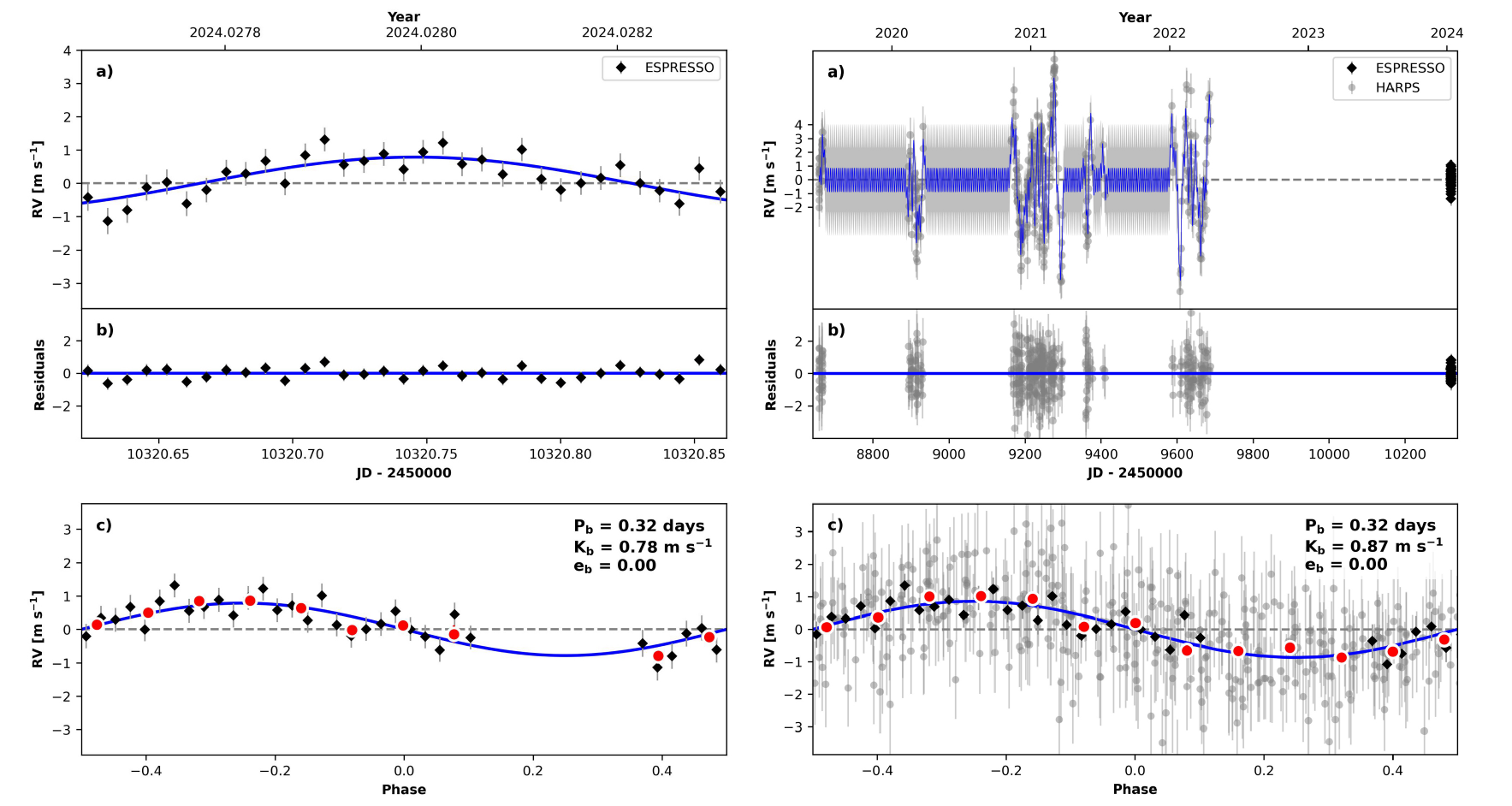}
    \caption{Radial Velocity (RV) variation of GJ~367 observed with \textbf{Left:} ESPRESSO (black diamonds) and \textbf{Right:} ESPRESSO + HARPS (grey circles). For both left and right plots, panel a) shows the time-seies RV measurements. The best-fit {\tt radvel} Gaussian Process (GP) single-Keplerian model for the ESPRESSO + HARPS combined dataset (grey line) used to remove stellar and instrumental correlated noise (see Sec.\ \ref{sec:rv}) is shown in the right hand side. The best-fit Keplerian orbital solution (blue line) is shown in each side. Panel b) shows the fit residuals (data - model). Panel c) shows the RV variations of GJ~367 as a function of orbital phase. The red points represent RVs binned to 10\% of $P_{\rm orb}$.}
    \label{fig:rvgp}
\end{figure*}

We explored the posterior distributions using {\tt radvel}'s implementation of {\tt emcee} \citep{emcee2013} to sample to the parameter space in a Markov Chain Monte Carlo (MCMC) scheme. We set 128 walkers for 10$^4$ runs in three ensembles for parallelization. The maximum auto-relative change (the stability of the autocorrelation time ($\tau$) for each parameter) achieved was 0.0148, much shorter than the length of a chain. The best-fit Keplerian solution for GJ~367~b is shown in the left panels of Fig.\ \ref{fig:rvgp}. Based on the ESPRESSO data alone, we find a RV semiamplitude of $K=0.78\pm0.12$ \ms\ corresponding to a planetary mass of $M_p=0.503\pm0.078$ M\e. This is in tension to only 1.4$\sigma$ with the reported mass measurement of 0.633$\pm$0.050 M\e\ from \citet{goffo_company_2023}, after quadratically combining the respective uncertainties given the two independent mass determinations. Using a multi-dimensional GP to account for stellar activity, \citet{goffo_company_2023} also found a mass of 0.546$\pm$0.093 M\e, which is in very good agreement with our ESPRESSO fit. A summary table of the various mass determinations from our work and the previous works is given in Table \ref{tab:masses}. 

\subsection{ESPRESSO + HARPS Data: Single Keplerian Gaussian Process Model}\label{ssec:gp}
After combining the 33 ESPRESSO and 371 HARPS RVs, we employed a Gaussian process (GP) regression model to disambiguate the planetary RV signal from stellar and instrumental correlated noise. We set the GP period such that long-period RV variations from the companion planets GJ~367~c and GJ~367~d were subsumed. We again used the {\tt radvel} package to construct the GP model and fit the Keplerian orbit. We chose the widely used quasi-periodic GP kernel \citep[e.g.,][]{Haywood2014,grunblatt2015,Dai2017} to model both the rotationally modulated stellar activity and stochastic instrumental drift plaguing the long-baseline combined dataset. A detailed description of how the quasiperiodic GP kernel takes form in the covariance matrix factored into the likelihood function is given in \citet{Dai2017}. The model basis was the same as in the Keplerian orbital fit above (orbital elements $P_{\rm orb},\,T_c,\,K,\,\sqrt{e}\cos{\omega},\,\sqrt{e}\sin{\omega}$). We imposed the same priors on $P_{\rm orb}$, $T_c$, $K$, $\sigma_{\rm ESPRESSO}$, $\gamma$, $\sqrt{e}\cos{\omega}$ and $\sqrt{e}\sin{\omega}$. A white noise jitter term for the HARPS RVs, $\sigma_{\rm HARPS}$, was added with a uniform prior. We also added the GP hyperparameters of variability amplitude $\eta_1$, non-periodic characteristic length $\eta_2$, variability period $\eta_3$, and periodic characteristic length $\eta_4$. Gaussian priors were placed on the hyperparameters based on estimations of stellar activity lifetimes from \tess.

We sampled the posterior distributions with {\tt emcee} in the same manner as above with a MCMC with 128 walkers for 10$^4$ runs in three ensembles. We show the best-fit GP and Keplerian orbit solution in the right-side panels of Fig. \ref{fig:rvgp}. The median values of the posteriors for the orbital parameters and GP hyperparameters are given in Table \ref{tab:params}. We find an RV semi-amplitude of $K=0.87\pm0.10$ \ms, corresponding to a planetary mass of $0.561\pm0.066$ M\e.

\subsection{ESPRESSO + HARPS Data: 3-Keplerian Gaussian Process Model}
We additionally fit a GP model including additional Keplerian terms for the known outer Super-Earth planets GJ 367 c and d \citep{goffo_company_2023}. The model basis included the same Keplerian orbital elements for each planet, and wide uniform priors were imposed based on the parameters derived in \citet{goffo_company_2023}. We placed broader priors on the GP hyperparameters, limiting the variability period to longer than the longest-period known planet ($\sim$34 days). We explored the posterior distributions with {\tt emcee} in the same manner as the other {\tt radvel} fits. From this amended GP fit with explicit Keplerian components for the additional planets, we find an RV semi-amplitude of $K=0.85_{-0.09}^{+0.08}$ \ms, corresponding to a planetary mass of $0.548\pm0.060$ M\e.

\begin{deluxetable*}{ccccc}[!ht]
\tablecaption{Mass measurements for GJ~367~b from literature and this work} 
\label{tab:masses}
\tablehead{
\colhead{Dataset} & \colhead{Fit Method} & \colhead{RV Semi-amplitude (\ms)} & \colhead{Planet Mass (M\e)} & \colhead{Reference}}
\startdata
ESPRESSO (N=33) & Keplerian & 0.78 $\pm$ 0.12 & 0.503 $\pm$ 0.078 $^{\bowtie}$ & This work\\
ESPRESSO (N=33) + HARPS (N=371) & GP (Single Keplerian) & 0.87 $\pm$ 0.10 & 0.561 $\pm$ 0.066 & This work\\
ESPRESSO (N=33) + HARPS (N=371) & GP (3-Keplerian) & 0.85$_{-0.09}^{+0.08}$ & 0.548 $\pm$ 0.060 & This work\\
ESPRESSO (N=33) + HARPS (N=275) & FCO & 0.81 $\pm$ 0.14 & 0.523 $\pm$ 0.091 & This work\\
\hline
HARPS (N=371) & GP & 0.86 $\pm$ 0.15 & 0.546 $\pm$ 0.093 & \citet{goffo_company_2023}\\
HARPS (N=275) & FCO (joint transit fit) & 1.003 $\pm$ 0.078 & 0.633 $\pm$ 0.050 $^{\bowtie}$ & \citet{goffo_company_2023}\\
HARPS (N=371) & Sines & 1.10 $\pm$ 0.14 & 0.699 $\pm$ 0.083 & \citet{goffo_company_2023}\\
\hline
HARPS (N=73) & FCO (joint transit fit) & 0.798 $\pm$ 0.11 & 0.546 $\pm$ 0.078 $^{\bowtie}$ & \citet{Lam2021}\\
HARPS (N=103) & GP & 0.87$^{+0.16}_{-0.15}$ & 0.559 $\pm$ 0.103 & \citet{Lam2021}\\
\enddata
\tablecomments{ $^{\bowtie}$= fiducial result. We elect to adopt the ESPRESSO-only Keplerian orbit fit results in this work over the others due to the advantage of single-night extreme-precision RV observations and simplicity of modeling. }
\end{deluxetable*}

\subsection{ESPRESSO + HARPS Data: Floating Chunk Offset Method}\label{ssec:fco}
As a check against overfitting from the GP regression, we used the Floating Chunk Offset (FCO) method \citep[e.g.,][]{Hatzes2010,Hatzes2011,Hatzes2014} to determine the RV semi-amplitude induced by GJ~367~b. The FCO technique enables the characterization of RV signals from USPs (P$_{\rm orb\,\lesssim}$  1 day) whose nightly induced Doppler reflex motion typically exceeds the RV variation caused by outer planets, stellar rotation, and night-to-night instrument instability \citep{Hatzes2019,Deeg2023}. An additive offset variable $\gamma_\mathrm{night}$ is assigned to nightly grouped ($N_\mathrm{obs}\ge 2$) RV measurements to focus on intranight RV variations while removing long-term ($P\ge 1$ day) systematics. 

RV measurements from single-measurement nights were excluded, leaving 275 HARPS RVs to combined with the ESPRESSO RVs. In our RV model we include the white noise component in the covariance function, and implicitly include the $\gamma_{night}$ terms in the measured RV values in the likelihood function. For sampling we again used {\tt emcee} to sample the posterior distributions with 512 walkers for 10,000 steps. We find $K=0.81\pm0.14$ \ms\ or 0.523$\pm$0.091 M\e, consistent with the results from the other two fits. This fit serves as a check against overfitting by the GP, but we do not adopt this result as it relies on a much larger number of parameters (each $\gamma_{night}$) than the other models.

\section{Tidal Distortion and Interior Composition}\label{ssec:distortion}
USPs experience extreme tidal forces due to their close proximity to their host stars. These forces can lead to significant tidal deformation, elongating the planet into an ellipsoidal shape, especially for planets that are fully molten. For GJ~367~b, its longer-period companion planets may excite its eccentricity, possibly inducing internal heating through tidal dissipation \citep{Jackson2010,correia2014}. This heating, combined with intense stellar irradiation, can raise surface temperatures well above 2000 K, potentially melting the surface rock and forming a permanent dayside magma ocean \citep[e.g.,][]{jackson2008,Jackson2010,Leger2011,Kite2016}. \citet{zhang_gj_2024} propose a permanently molten dayside for GJ~367~b. In such environments, surface materials may vaporize and form mineral-rich atmospheres, which can be stripped away by stellar winds or condense on the nightside, leading to rock vapor cycles akin to weather \citep{Schaefer2009,Leger2011,demory_map_2016,Kite2016}. These effects may be especially pronounced for GJ~367~b due to its lower gravity, making it easier for molten material or vapor to escape.

\subsection{First-Order Tidal Distortion: Love's Theory}
From our transit radius of $0.736\pm0.035$ R\e\ and mass of $0.503\pm0.078$ M\e, we find a bulk density of $\rho_p = 6.9^{+1.6}_{-1.4}$ \gcm\ (assuming a spherical planet). From this, the Roche limit of GJ~367~b corresponds to $P_{\mathrm{Roche}} \approx 4.8$ hours, yielding $P_{\mathrm{orb}}/P_{\mathrm{Roche}} \approx 1.61 \pm 0.16$ under the assumption of an incompressible fluid body \citep{rappaport_roche_2013}. This ratio indicates that GJ~367~b is not as perilously close to tidal disruption as some other USPs \citep[e.g., KOI-1843.03, TOI-6255~b, TOI-6324~b, and TOI-2431~b;][]{rappaport_roche_2013,Dai2024,Lee2025,Tas2025}, but its compact orbit and possible interactions with outer planets still place it in a regime where tidal effects are non-negligible. Based on the constant-lag-angle model of equilibrium tides for slowly rotating stars \citep{Goldreich1966,winn_kepler-78_2018,Dai2024}, the tidal decay timescale is described by star-to-planet mass ratio, stellar density, present-day orbital period, and a nominal stellar tidal quality factor (assume $Q^{\prime}_\star=10^7$ for a mature M dwarf). We find a tidal decay timescale of $\tau_p\approx26$ Gyr, and GJ~367~b would reach the Roche limit within $\tau_{\rm Roche}\approx3$ Gyr. Eccentricity tides may influence the distortion of GJ~367~b, as has been observed on Io \citep{Peale1979}, one of the most tidally deformed objects in our Solar System. If we consider tidal dissipation due to a non-zero eccentricity (which could be excited by the longer-period planets in the system), the tidal decay timescale is instead $\tau_{P,ecc}\approx2$ Gyr (see Eq. 14 of \citealt{Dai2024}), assuming a planetary tidal quality factor $Q^\prime_p=1000$ and $e=0.001$ \citep[Io's eccentricity is $\sim0.004$;][]{Peale1979}. 

The amplitude of the radial tidal displacement ($\delta R_p$) for a solid, homogeneous planet can be estimated using the second-order Love number, $h_2$ \citep{Love1944}, which encodes the elastic response of a body to external tidal forcing. This is described by the expression $h_2=\frac{5/2}{1+\tilde{\mu}}$, where $\tilde{\mu}=\frac{19\mu}{2\rho gR_p}$. The terms $\mu$, $\rho$, and $g$ are the tensile strength, mean density, and surface gravity of the planet, respectively. For Earth, $h_2$ lies in the range $0.6$–$0.9$ \citep{lambeck1980}; let us assume $h_2 = 2.5$ for GJ~367~b, corresponding to fully molten planet ($\mu$=0) due to extreme internal heating from secular interactions. Under this assumption, we estimate $\delta R_p\approx0.048$. The mean volumetric radius of the planet $R_{\mathrm{vol}}$ describes the triaxial radii of the ellipsoidal planet as $R_{\mathrm{vol}}\equiv(R_1R_2R_3)^{1/3}$, where $R_1$ points along the star-planet axis, $R_2$ along the orbital axis, and $R_3$ along the polar axis. From the relation $R_{\mathrm{vol}} = R_{\mathrm{transit}}(1 + \tfrac{7}{12}\delta R_p)$ \citep{Dai2024}, this correction yields $R_{\mathrm{vol}}/R_{\mathrm{transit}} \approx 1.027$, or a modest $\sim$2.7\% increase in the radius. We compare $R_{\mathrm{transit}}$ and the Love model $R_{\mathrm{vol}}$ in the left panel of Fig.~\ref{fig:mrd}. This estimate of distortion does not qualitatively change our conclusions about the interior composition of GJ~367~b according to the Preliminary Reference Earth Model \citep[PREM;][]{PREM} mass-radius relation of \citet{zeng2016}. 

While this simple estimate provides an illustrative first-order correction, it likely underestimates the true degree of distortion. In reality, the tidal response depends on the planet’s layered structure, rheology, and melting, none of which are captured by a homogeneous Love number approximation. To more rigorously assess the magnitude of tidal distortion for GJ~367~b, we turn to a fully self-consistent tidal distortion model in the next subsection.

\subsection{Self-Consistent Tidal Distortion and Interior Composition Model}
To better constrain the possible tidal distortion of GJ~367~b, we apply the \texttt{greenlantern} code \citep{Price+2025} to the JWST white lightcurve transit observations from \citet{zhang_gj_2024}. We use \texttt{celerite} \citep{ForemanMackey+2017} to evaluate the log-likelihood of the residual between the observed data and a trial model for given correlated ``red'' noise parameters. For an initial and efficient exploration of the full parameter space, we apply \texttt{pocoMC} \citep{pocomc1,pocomc2} in conjunction with the neural autoregressive flow \citep[NAF,][]{Huang+2018} provided by \texttt{zuko} \citep{zuko}. After this initial fitting, we use \texttt{emcee} \citep{emcee2013} to more carefully explore around the maximum likelihood point, where the estimated standard deviation determines the initial spread of the MCMC walkers.

We next apply the \texttt{sisyphus} code (Price et al., in prep) to model possible interior structures of GJ~367~b for selected core and mantle materials with fixed transit radius. This code uses a relaxation method, based on \citet{Price2020}, to self-consistently determine the density structure of a tidally-locked planet, given a stellar mass, orbital period, planet materials, and constraints on the transit radius and core-mass fraction; taking into account the stellar gravitational field, centrifugal forces from rapid orbital motion, and planet self-gravity; and assuming that the system is in equilibrium, so the gravitational, rotational, and hydrostatic forces balance. We consider several possible materials for the core and mantle of the planet: We take the equations of state for alpha iron ($\alpha$-\ce{Fe}), iron sulfide (\ce{FeS}), perovskite (\ce{MgSiO3}), and enstatite (\ce{MgSiO3}) from the collation of \citet[see references therein]{Seager2007}; for an Earth-like core and mantle from \citet{zeng2016}; and for pyrolite magma from \citet{Boley+2023}. We assume that the transition pressure from enstatite to perovskite is 23~GPa, following \citet{Sotin+2007}. We consider four possible combinations of these materials in constructing a two-layer planet: an alpha iron core surrounded by a perovskite and enstatite mantle (Fe--pv--en); an iron sulfide core surrounded by a perovskite and enstatite mantle (FeS--pv--en); an iron core surrounded by magma (Fe--melt); and the PREM core and mantle \citep{zeng2016}. For fixed transit radius, composition, stellar mass, and orbital period, we sweep through values of core mass fraction (CMF), optimizing to find the planet density structure that meets the constraints without being tidally disrupted. Unlike \citet{Price2020}, we use a novel, doubly-iterative procedure that employs a rootfinder in conjunction with the self-consistent field method \citep{Hachisu1986a,Hachisu1986b,Price2020}; this allows us to perform more targeted simulations of planets in a given region of parameter space, avoiding interpolation between models altogether.

\begin{figure*}[!ht]
    \begin{center}
    \includegraphics[angle=270,width=0.49\linewidth]{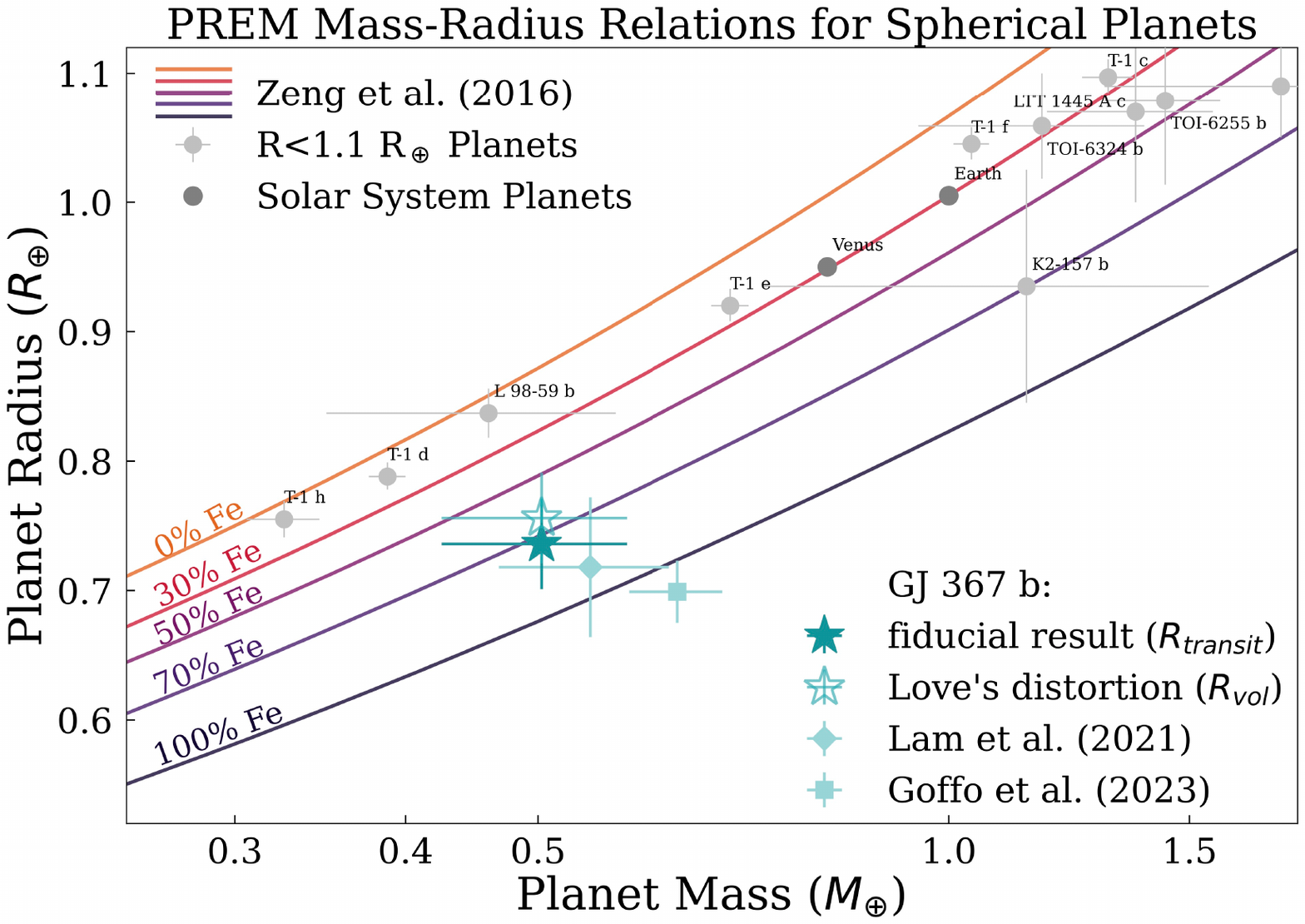}
    \includegraphics[angle=270,width=0.49\linewidth]{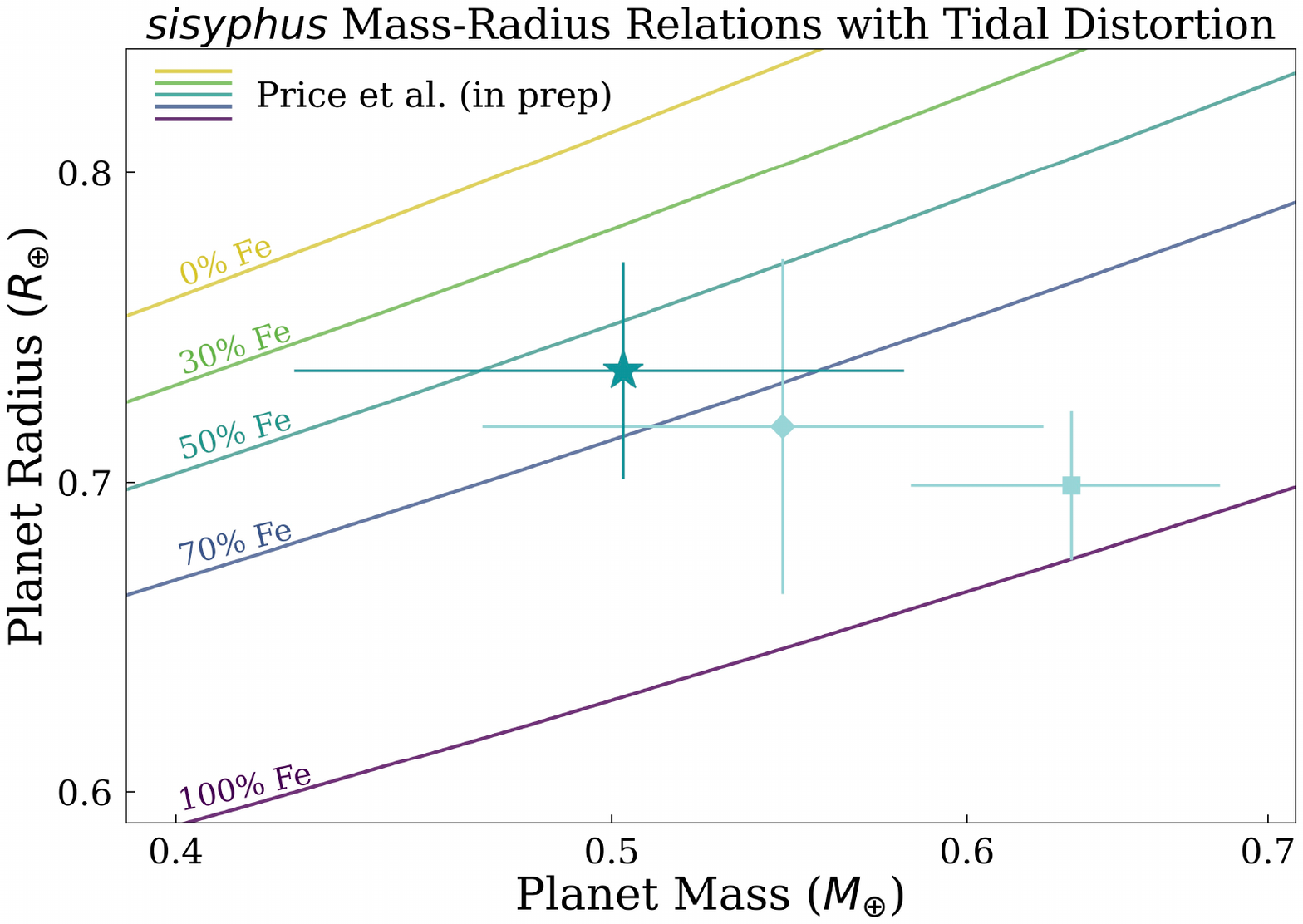}
    \caption{\textbf{Left:} Mass-radius diagram of exoplanets with $<35$\% mass measurement uncertainty, taken from the NASA Exoplanet Archive\textsuperscript{a}. The adopted RV mass and transit radius measurements of GJ~367~b from this work (teal star) are shown with those from previous works (light blue points). The teal outlined star represents the volumetric radius if we assume tidal distortion according to Love's theory (see Sec.\ \ref{ssec:distortion}). The model curves are based on the Preliminary Reference Earth Model \citep[PREM;][]{PREM} composition, which is based on the equation of state extrapolated from Earth's seismic density profile \citep{zeng2016}. \textbf{Right:} Modified mass-radius relations for a theoretical planet orbiting GJ~367 with the same orbital period as GJ~367~b, assuming an iron core and perovskite + enstatite mantle composition (see Sec.\ \ref{ssec:distortion}). The plotted points are the same as in the Left-hand panel. When tidal distortion is taken into account in the mass-radius relation, the measured mass and radius from \citet{goffo_company_2023} does not imply a planet more dense than iron; assuming a spherical shape in the mass-radius relation for a distorted planet underestimates the volume, leading to an overestimate of bulk density as in the left-hand panel. The mass and radius measured in this work place GJ~367~b between the 50\% and 70\% iron mass fraction curves, consistent with a composition less iron-rich than Mercury \citep{Hauck+2013}.}
    \end{center}
    \small\textsuperscript{a} \url{https://exoplanetarchive.ipac.caltech.edu/}
    \label{fig:mrd}
\end{figure*}

\begin{figure}
    \centering
        \includegraphics[width=\linewidth]{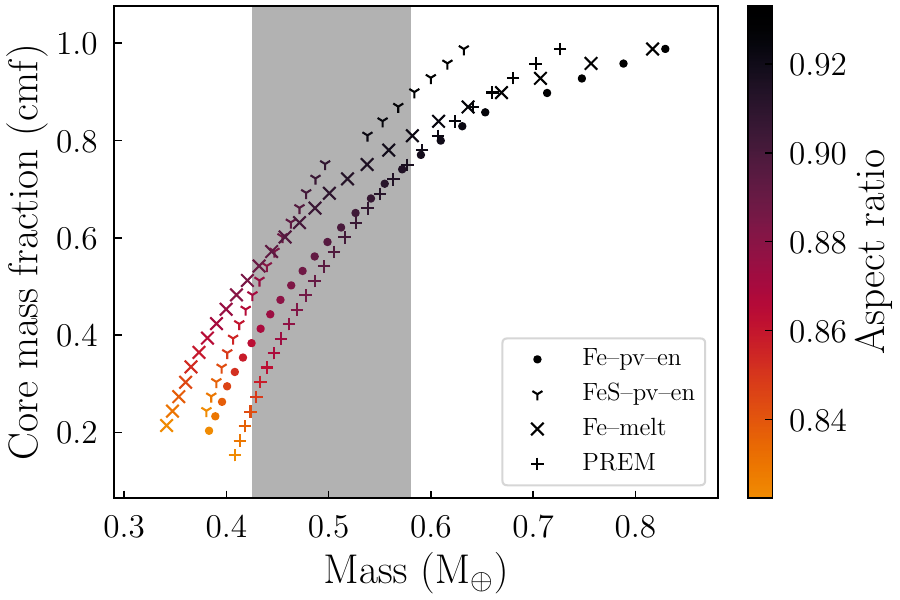}
    \includegraphics[width=\linewidth]{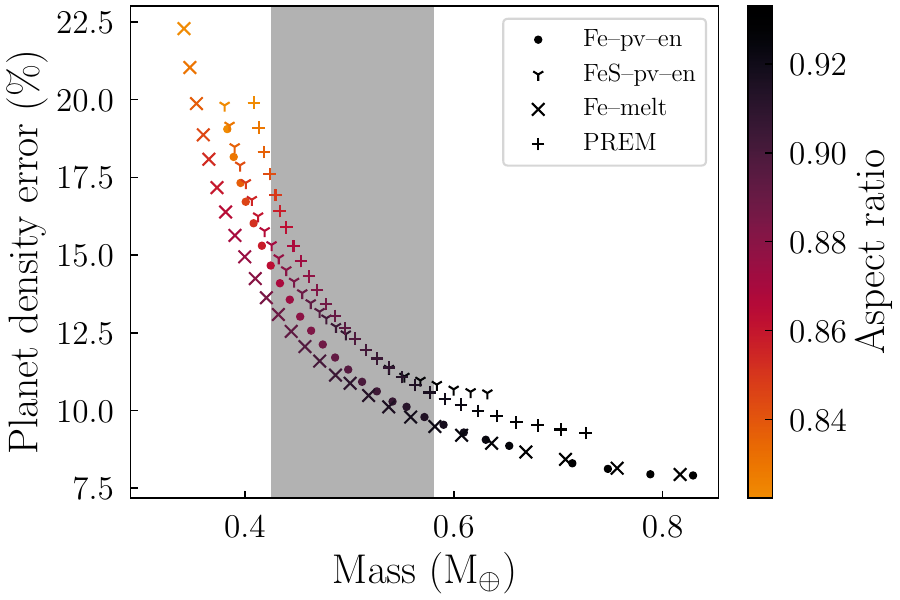}
    \caption{\textbf{Top:}  Mass vs. core mass fraction (CMF) and \textbf{Bottom:} mass vs. planet density error as a function of radial aspect ratio (color bars) from tidal distortion. We show the measured mass of GJ 367 b ($0.503 \pm 0.078$~M\e) as the grey shaded region. We adopt the Fe--pv--en (iron core surrounded by perovskite-enstatite mantle) composition in the right panel of Fig.\ \ref{fig:mrd}.}
    \label{fig:cmf}
\end{figure}

In the right-side panel of Figure~\ref{fig:mrd}, we provide modified mass-radius relations, tailored to GJ~367~b for a single material combination (Fe--pv--en), across several iron mass fractions, and accounting for tidal distortion. The \citet{goffo_company_2023} and \citet{Lam2021} results, as well as the one reported in this work, fall above the mass-radius curve of a theoretical 100\% iron planet if tidal distortion is considered. The mass and mean radius reported above are consistent with a bulk composition between 50\% and 70\% iron by mass; for comparison, Mercury is about 70\% iron by mass \citep{Hauck+2013}. Considering all material composition scenarios and a range of aspect ratios due to tidal distortion, we find the CMF is never consistent with a pure iron planet (Fig.~\ref{fig:cmf}).

\section{Discussion} \label{sec:discussion}

When characterizing planets around M dwarfs, it is essential to account for observational and astrophysical effects that can bias radius measurements, since even small errors in planetary radius propagate into large uncertainties in interior composition models. As we discussed in Sec.\ \ref{ssec:dilution}, a key observational challenge is photometric contamination from unresolved nearby stars, which dilutes the transit depth and leads to underestimated planetary sizes \citep[see Fig.\ \ref{fig:contamination};][]{han_tessgaia_2023,Han2025}. Additionally, the geometry of the planet itself must be considered, as USPs may be significantly tidally distorted \citep[see Sec.\ \ref{ssec:distortion}; e.g.,][]{zhang2024,Dai2024,Lee2025}. Failing to model this effect leads to an underestimation of radius and overestimation of bulk density. Finally, the strong surface magnetic activity of M dwarfs introduces two systematic biases: (1) magnetic inhibition of convection can inflate stellar radii \citep[e.g.,][]{Wanderley2023}, and (2) non-uniform surface brightness from spots and faculae can distort transit depths \citep[e.g.,][]{Mori2024}. In the following subsections, we examine these caveats in detail and assess their impact on the derived radius and composition of GJ~367~b.

\begin{deluxetable*}{lll}
\tablecaption{Systematics affecting planetary radius measurements.} 
\label{tab:bias}
\tablehead{
\colhead{Mechanism} & \colhead{Observational Bias} & \colhead{Possible Effect on our $R_p$ Measurement}
}
\startdata
Photometric contamination & Dilutes the transit depth & $\lesssim$1\% underestimate \\
from nearby stars & & \\
\tableline
Photometric variability & Star spots may bias the baseline flux, & $\lesssim$1\% underestimate \\
due to surface activity & or spot crossings during transit may & \citep{Nichols-Fleming2020} \\
& dilute the transit signal & \\
\tableline
Empirical scatter in M dwarf & Overestimated $R_\star$, & $\lesssim$5\% underestimate \\
radii of the same mass and & underestimated $R_p$ & \\
$T_{\rm eff}$ \citep{Mann2015} & & \\
\tableline
M dwarf radii are underestimated & Underestimated $R_\star$, & $\lesssim$5\% overestimate \\
by MIST and other evolutionary & overestimated $R_p$ & \\
isochrone models compared to & & \\
magnetic models (e.g., SPOTS) & & \\
\tableline
Tidal distortion of the planet & Planet-star radius longer & $\lesssim$10\% difference between \\
& than polar radius & transit $R_{\rm transit}$ and $R_{\rm vol}$ (Fig.\ \ref{fig:cmf}) \\
\enddata
\end{deluxetable*}

\subsection{Stellar Magnetic Field}\label{ssec:mag}
The magnetic fields of M dwarfs drive substantial starspot and faculae coverage, introducing rotationally modulated photometric variability and spectral line distortions that may bias both transit depth measurements \citep{Rackham2018, Rackham2019,Bustos2025} and radial velocity signals \citep{Reiners2010, Robertson2014}. Moreover, magnetic stellar structure models \citep[e.g., SPOTS, Dartmouth models;][]{Somers2020,Feiden2012,Feiden2015,Feiden2016} predict inflated stellar radii and reduced effective temperatures compared to non-magnetic models \citep[e.g., MESA Isochrones and Stellar Tracks (MIST);][]{MIST}, leading to systematic offsets in derived planetary radii and equilibrium temperatures \citep{Kochukhov2021}. Such discrepancies bias derived densities and compositions and complicate the interpretation of planetary atmosphere signals. Additionally, M dwarfs often exhibit strong chromospheric and coronal activity, generating ultraviolet and X-ray radiation that can drive atmospheric escape on close-in planets \citep{Luger2015, Loyd2018}, further complicating inferences about their present-day atmospheres.

Young M dwarfs ($<$ 1 Gyr) exhibit strong surface magnetic activity, with global magnetic field strengths typically orders of magnitude stronger than in Sun-like stars \citep[][]{Kochukhov2021}. \citet{goffo_company_2023} measured the mean longitudinal magnetic field of GJ~367 to be $\langle B_z\rangle=-7.3\pm3.2$ G, which is considerably lower than typical mean longitudinal field strengths in young, active M dwarfs \citep[100s of G;][]{donati2008}. If GJ~367 is indeed a quiet star as \citet{goffo_company_2023} suggest, this longitudinal field measurement should be representative of the global field. However, if the field is predominantly non-antisymmetric, the longitudinal field strength would not characterize the global field. Although GJ~367 is not likely to be very active based on its longitudinal magnetic field strength and older age \citep[$\sim$4-8 Gyr;][]{goffo_company_2023}, a full magnetic field analysis of the star is necessary to draw any firm conclusions, and possible measurement biases affecting our estimate of $R_p$ remain.

Depending on a planet's transit chord, non-uniform surface brightness caused by spots may bias the transit depth measurement. Zeeman Doppler Imaging (ZDI), combined with Principal Component Analysis (PCA) provide some of the only insight into the magnetic field morphologies and spot distributions on M dwarfs \citep[e.g.,][]{Lehmann2024}, especially for stars with relatively simple morphologies and moderately projected equatorial velocities. \citet{See2025} compiled an extensive catalog of ZDI data for 96 M dwarfs, however, generalizations of M dwarf field strengths and morphologies do not necessarily hold across masses, temperatures, ages, and Rossby number. \citet{Lehmann2024} constructed a ZDI map of Gl~617B, a $\sim$ 0.45 M\sol\ M dwarf with a $P_{\rm rot}$ of $\sim$40 days, somewhat comparable to GJ~367 ($\sim$0.45 M\sol, $P_{rot}\approx$50 days). The magnetic field of GJ~617B is predominantly poloidal (98\%) and axisymmetric (98\% of the magnetic energy is contained in $m$=0 modes only). The surface averaged unsigned magnetic field strength ranges from $\langle B_v\rangle=36-75$ G across three observing seasons, with a toroidal field strength on the order $\langle B_{tor}\rangle\approx5$ G. If GJ~367 has similar magnetic field characteristics, we would expect surface brightness variations to minimally affect the transit measurement. If there are significant surface brightness and temperature variations across the disk of the star caused by the magnetic field, multi-band transit and stellar activity modeling is required to accurately characterize the transit \citep[e.g.,][]{Mori2024}. Generally, slow rotators ($\sim$50 day) such as GJ 367 are advantageous: reduced spot coverage and variability, as well as reduced flaring, lead to more stable baselines and lower stellar contamination \citep[$<$1\%;][]{Nichols-Fleming2020}. However, \citet{Lehmann2024} find that among their sample, some slowly rotating M dwarfs can still host strong magnetic fields. A detailed study of the magnetic field of GJ~367 is required to reveal the degree to which non-uniform surface brightness may affect the transit depth measurement. 

It is not yet fully understood how the magnetic fields and rotation of M dwarfs affect their radii, thus complicating their characterization. Radius inflation, a notorious problem in M dwarf characterization, is generally accepted to be dependent on rotation, with the radii of slower rotators tending to match theoretical predictions \citep{Parsons2018}. However, there is a large scatter in rotation rates for M dwarfs of the same age, mass, and T$_{\rm eff}$ \citep{Mann2015,Nichols-Fleming2020}. Furthermore, different evolutionary models (e.g., MIST, DARTMOUTH, PARSEC, SPOTS) predict different radii for M dwarfs of the same T$_{\rm eff}$, even for slow rotators \citep{Wanderley2023}, by as much as $\sim$15\%. Thus, planet radius measurements in M dwarf systems remain uncertain until stellar radius modeling improves. A summary of the approximate contributions to the measurement error of $R_p$ by these various effects are given in Table \ref{tab:bias}.


\subsection{Possible Formation History of GJ~367~b}\label{ssec:formation}

Although we find a lower bulk density for GJ~367~b than previously established, it still appears to have an anomalously enhanced iron fraction compared to other confirmed sub-Earths (e.g., TRAPPIST-1 planets, see Fig. \ref{fig:mrd}). The similarly high core-mass fraction (CMF) of Mercury has long motivated a variety of formation hypotheses, including large-scale mantle stripping via one or more giant impacts \citep{Benz1988,Benz2007}, preferential removal of silicates through high-temperature vaporization and volatile loss \citep{Lupu2014}, and aerodynamic fractionation of metal and silicate grains in the protoplanetary disk \citep{Weidenschilling1978}. Alternative models invoke photophoretic forces to drive radial separation of metals and silicates in the hot inner nebula, producing inherently metal-rich planetesimals \citep{Wurm2013}. Geochemical analyses of Mercury’s surface and meteoritic analogs suggest that magnetic, thermal, and chemical processing in the early disk may also have played a role \citep{Ebel2017}. These pathways—each potentially operating under different stellar and disk conditions—provide a physical framework for interpreting the origin of super-Mercury exoplanets, whose high densities may reflect similar dynamical and thermochemical histories. In particular, the proximity of planets like GJ~367~b to magnetically active M-dwarf hosts may enhance some of these processes, making them compelling laboratories for testing Mercury-like formation scenarios in extreme environments. 

In the framework of giant impacts, high-velocity collisions between differentiated protoplanets can preferentially remove silicate mantles while leaving iron cores largely intact, thereby increasing the bulk density of the resulting body \citep{Marcus2010}. Numerical simulations demonstrate that while such events can produce planets with CMFs approaching or even exceeding that of Mercury, they are relatively rare outcomes, requiring finely tuned impact parameters such as high velocities, grazing geometries, or specific mass ratios between target and impactor \citep{Marcus2010}. However, a critical challenge lies in preventing the stripped silicate debris from re-accreting onto the surviving core, which would erase the compositional signature of mantle loss. Mechanisms such as dynamical ejection of debris, aerodynamic drag causing silicate vapor dispersal, or radiative blow-off in the high-energy environments of young stars have been proposed to suppress re-accretion \citep[e.g.,][]{Benz2007,Asphaug2014,Kite2016}, but the efficiency of these processes remains uncertain. Thus, although maximum collisional stripping provides a viable mechanism for generating high-CMF planets, accounting for the most extreme cases—where inferred densities rival or surpass that of pure iron—likely demands the involvement of additional processes, or a synergy between impact dynamics and early environmental conditions.

Recent N-body simulations with realistic collision physics suggest that such dense planets can arise in dynamically excited inner disks perturbed by giant planets \citep{Scora2024}. In these models, Mercury analogues with CMFs > 0.4 form in roughly 10\% of cases when additional embryos are seeded interior to 0.7 AU, though most survivors have CMFs below Mercury’s value owing to subsequent accretion of silicate-rich debris. The likelihood of producing super-dense analogues increases in high-excitation environments or when the initial mass distribution is concentrated toward the star, leading to frequent high-energy impacts and incomplete mantle re-accretion. This framework naturally extends to super-Mercury USPs. Their high densities and small sizes, exemplified by GJ~367~b, may similarly record a history of energetic collisional erosion, selective debris loss, and suppressed re-accretion within compact, magnetically active planetary systems. In M dwarf environments, where stellar magnetic fields and activity complicate planet characterization \citep{Kochukhov2021,Somers2020}, formation pathways involving excited early dynamics offer a promising route to explain the observed diversity in bulk compositions.

\section{Summary}\label{sec:sum}
We report revised mass and radius measurements for GJ~367~b based on single-night full-orbit RV observations with ESPRESSO and the latest \tess\ photometry, and discuss possible observational and modeling systematics affecting these measurements. We find that while GJ~367~b is indeed iron-rich, as reported in the \citet{Lam2021} discovery paper, its mass and radius are not consistent with an interior comparable to pure iron as suggested in \citet{goffo_company_2023}. If we account for tidal deformation at GJ 367 b, the iron fraction of the planet is $\sim$50-70\%, more consistent with our own Mercury \citep{Hauck+2013}. This reduced CMF and iron fraction is in better alignment with giant impact simulations to explain the formation of iron-rich rocky planets.

\nolinenumbers

\vspace{0.5cm} 
\noindent {\bf ACKNOWLEDGEMENTS} 

R.A.L. acknowledges this material is based upon work supported by the National Science Foundation Graduate Research Fellowship Program under grant No. 1842402 and grant No. 2236415. Any opinions, findings, and conclusions or recommendations expressed in this material are those of the author(s) and do not necessarily reflect the views of the National Science Foundation. 
    
E.M.P. and M.Z. gratefully acknowledge the generous support of the Heising-Simons Foundation through the 51 Pegasi b Postdoctoral Fellowship.

This material is based upon work supported by the National Aeronautics and Space Administration under Grant No. 80NSSC25K7159 issued through the Exoplanets Research Program (XRP).
    
We thank Jon Jenkins and Joseph Twicken for helpful discussions of the SPOC pipeline contamination factor estimates and dilution correction.

We acknowledge the use of public TESS data from pipelines at the TESS Science Office and at the TESS Science Processing Operations Center. Resources supporting this work were provided by the NASA HighEnd Computing (HEC) Program through the NASA Advanced Supercomputing (NAS) Division at Ames Research Center for the production of the SPOC data products.

This article made use of data collected by the TESS mission and are publicly available from the Mikulski Archive for Space Telescopes (MAST) operated by the Space Telescope Science Institute (STScI). Funding for the TESS mission is provided by NASA’s Science Mission Directorate. K.A.C. and C.N.W. acknowledge support from the TESS mission via subaward s3449 from MIT.

    \software{\texttt{celerite} \citep{ForemanMackey+2017}, \texttt{emcee} \citep{emcee2013}, \texttt{pocoMC} \citep{pocomc2}, \tt{RadVel} \citep{fulton_radvel_2018}, \tt{exoplanet} \citep{foreman-mackey_exoplanet_2021}}

    \textit{Data:} The SPOC TESS data used in this work can be found in MAST: \dataset[10.17909/t9-nmc8-f686]{http://dx.doi.org/10.17909/t9-nmc8-f686}. This work uses data supplied from the NASA Exoplanet Archive: \dataset[10.26134/ExoFOP5]{http://dx.doi.org/10.26134/ExoFOP5}.

    \facilities{Very Large Telescope, Paranal Observatory, \tess}

\pagebreak

\bibliography{bib}
\bibliographystyle{aasjournal}

\pagebreak

\appendix

We compared the sector-wise contamination factors computed from both SPOC and TGLC. The contamination factor is from both pipelines is simply the total flux from nearby stars divided by the total flux from the target star within the pipeline's aperture. The TGLC pipeline uses a fixed 3x3 pixel aperture while the SPOC PDCSAP aperture ($\sim$16-20 pixels for GJ 367) is dynamically determined based on crowding and instrumental systematics. This difference in apertures may account for the differences in the contamination estimates.

\begin{figure}
    \centering
    \includegraphics[width=0.5\linewidth]{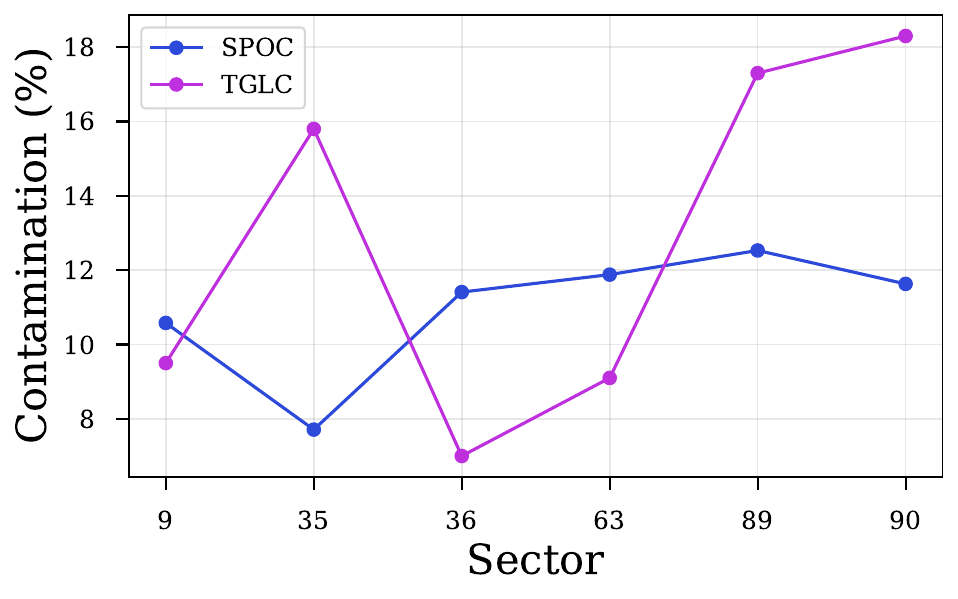}
    \caption{Contamination factor (\%) from SPOC PDCSAP and TGLC light curves for each \tess\ sector.  }
    \label{fig:contamination}
\end{figure}

\begin{deluxetable*}{lccCC}\label{tab:dist_fit}
    \tablecolumns{5}
    \tablehead{\colhead{Parameter} & \colhead{Symbol} & \colhead{Prior\tablenotemark{\scriptsize a}} & \colhead{Median and error} & \colhead{Maximum likelihood value}}
    \tablecaption{Prior distributions used in and derived values from Bayesian lightcurve fitting. \label{tbl:lc}}
    \startdata
        Mean radius ratio\tablenotemark{\scriptsize b} & $r$ & $\log\mathcal{U}\!\left[10^{-3}, 10^{-1}\right]$ & 0.014447 \substack{+0.0010408 \\ -0.0011024} & 0.014398 \\
        Flattening & $f$ & KDE\tablenotemark{\scriptsize c} & 0.10555 \substack{+0.043973 \\ -0.025167} & 0.079827 \\
        Scaled semimajor axis\tablenotemark{\scriptsize d} & $a / R_\star$ & $\mathcal{N}\!\left[a_0 / R_\star, 0.1\right]$ & 3.3195 \substack{+0.10157 \\ -0.098817} & 3.2963 \\
        Mid-transit time offset\tablenotemark{\scriptsize e} (days) & $t_0$ & $\mathcal{N}\!\left[t_{0,\text{est}}, 0.01\right]$ & -0.0016823 \substack{+0.00010865 \\ -0.00012814} & -0.0016747 \\
        Impact parameter & $b$ & $\mathcal{U}\!\left[0, 1\right]$ & 0.56347 \substack{+0.040102 \\ -0.046911} & 0.57581 \\
        Limb darkening coefficients\tablenotemark{\scriptsize f} & $q_1$ & $\mathcal{U}\!\left[0,1\right]$ & 0.12180 \substack{+0.18636 \\ -0.09746} & 0.0087329 \\
        & $q_2$ & $\mathcal{U}\!\left[0,1\right]$ & 0.45792 \substack{+0.35964 \\ -0.32072} & 0.47140 \\
        Mat\'ern-3/2 amplitude\tablenotemark{\scriptsize g,h} & $\sigma_{M}$ & $\log\mathcal{U}\!\left[10^{-2} \sigma, 10^2 \sigma\right]$ & -9.7141 \substack{+0.079875 \\ -0.078760} & -9.7369 \\
        Mat\'ern-3/2 timescale\tablenotemark{\scriptsize g,i} & $\rho_{M}$ & $\log\mathcal{U}\!\left[10^{-2} \tau, 10^2 \tau\right]$ & -5.7185 \substack{+0.033097 \\ -0.015102} & -5.7335 \\
        Jitter amplitude\tablenotemark{\scriptsize g,h} & $\sigma_{J}$ & $\log\mathcal{U}\!\left[10^{-2} \sigma, 10^2 \sigma\right]$ & -8.3849 \substack{+0.0035113 \\ -0.0035089} & -8.3841
    \enddata
    \tablecomments{\tablenotetext{\scriptstyle a}{We use $\mathcal{U}\!\left[a, b\right]$ to indicate the uniform distribution on the interval $\left(a, b\right)$; $\log\mathcal{U}\!\left[a, b\right]$ to indicate the log uniform, or reciprocal, distribution on the interval $\left(a, b\right)$; and $\mathcal{N}\!\left[\mu, \sigma\right]$ to indicate the normal distribution of mean $\mu$ and standard deviation $\sigma$.}
    \tablenotetext{\scriptstyle b}{The mean radius ratio is defined as the geometric mean $r \equiv \left(R_a R_b R_c\right)^{1/3} \big/ R_\star$. We assume furthermore that $R_b = R_c$.}
    \tablenotetext{\scriptstyle c}{The prior on flattening is informed by a kernel density estimate based on interior structure modeling, assuming the composition components Fe--pv--en and the planet mass measured in this work.}
    \tablenotetext{\scriptstyle d}{The JWST white lightcurve includes a single transit of GJ~367~b, so the orbital period cannot be constrained from the JWST transit data alone. Therefore, we adopt the value $P_\text{orb} = 0.3219225$~days from \citet{goffo_company_2023}. Combined with the estimated stellar density $\rho_{\star,\text{est}} = 6.646$~g~cm${}^{-3}$, consistent with Table~\ref{tab:params}, the orbital period uniquely sets $a_0 / R_\star$ via $a_0 / R_\star = \sqrt[3]{G \rho^{}_{\star,\text{est}} P_\text{orb}^2 \big/ 3 \pi}$.}
    \tablenotetext{\scriptstyle e}{The mid-transit time offset mean was estimated by eye and given an artificially inflated standard deviation.}
    \tablenotetext{\scriptstyle f}{We adopt the quadratic limb darkening parameterization of \citet{exoplanet:kipping13}, which derives a nonlinear mapping from $q_1, q_2$ to the standard coefficients $u_1, u_2$.}
    \tablenotetext{\scriptstyle g}{The reported value is the natural logarithm.}
    \tablenotetext{\scriptstyle h}{The parameter $\sigma$ here refers to the estimated out-of-transit standard deviation.}
    \tablenotetext{\scriptstyle i}{The parameter $\tau$ here refers to an estimated characteristic timescale of $P_\text{orb}$.}}
    
\end{deluxetable*}




\end{document}